\documentclass[10pt,twocolumn,letterpaper]{article}
\usepackage[final,algorithms]{wacv_2025} 

\usepackage{graphicx}
\usepackage{amsmath}
\usepackage{amssymb}
\usepackage{booktabs}

\usepackage[dvipsnames]{xcolor}       %

\definecolor{citecolor}{HTML}{0071BC}
\definecolor{linkcolor}{HTML}{ED1C24}
\usepackage[pagebackref,breaklinks,colorlinks,citecolor=citecolor, linkcolor=linkcolor]{hyperref}

\usepackage{orcidlink}

\usepackage{comment}

\usepackage{graphicx}
\usepackage{subcaption}
\usepackage{float}

\usepackage{booktabs}                   
\usepackage{multirow}
\usepackage{array}
\usepackage{paralist}
\usepackage{enumitem}

\usepackage{mathtools}
\usepackage{amssymb,amsmath}   
\usepackage{color}

\usepackage{pifont}
\usepackage{xspace}

\usepackage{diagbox}

\usepackage[export]{adjustbox}
\usepackage{sidecap}

\usepackage{multirow, multicol, makecell, booktabs}
\usepackage{wrapfig}

\def\eg{e.g.,~}               
\def\ie{i.e.,~}               

\newcommand{\figref}[1]{Figure~\ref{fig:#1}} 
\newcommand{\tabref}[1]{Table~\ref{tab:#1}}
\newcommand{\eqnref}[1]{Eq.~\ref{eq:#1}}

\long\def\ignorethis#1{}

\definecolor{darkmagenta}{rgb}{0.55, 0.0, 0.55}
\definecolor{lightpastelpurple}{rgb}{0.69, 0.61, 0.85}

\newcommand{\hcolor}[1]{\textcolor{blue}{\textbf{#1}}}

\newcommand{\ocolor}[1]{\textcolor{magenta}{\textbf{#1}}}

\usepackage{enumitem}
\setlist{topsep=-1pt,itemsep=-1.5pt,leftmargin=5mm}

\newlength\paramargin
\newlength\figmargin
\newlength\subfigmargin
\newlength\secmargin
\newlength\subsecmargin
\newlength\tabmargin
\newlength\eqmargin
\newlength\tabupmargin

\newcolumntype{C}[1]{>{\centering\let\newline\\\arraybackslash\hspace{0pt}}m{#1}}

\newcommand*\colourmark[1]{%
  \expandafter\newcommand\csname #1xmark\endcsname{\textcolor{#1}{\ding{56}}}%
}

\colourmark{blue}
\colourmark{red}

\newcommand*\colourchecksnow[1]{%
  \expandafter\newcommand\csname #1snow\endcsname{\textcolor{#1}{\ding{100}}}%
}

\colourchecksnow{cyan}

\definecolor{darkspringgreen}{rgb}{0, 0.6, 0.3}
\newcommand*\colourcheck[1]{%
  \expandafter\newcommand\csname #1check\endcsname{\textcolor{#1}{\ding{52}}}%
}

\colourcheck{blue}
\colourcheck{cadmiumgreen}

\newcommand*\colourtri[1]{%
  \expandafter\newcommand\csname #1tri\endcsname{\textcolor{#1}{\ding{115}}}%
}

\colourtri{black}
\colourtri{red}

\makeatletter
\@namedef{ver@everyshi.sty}{}
\makeatother
\usepackage{tikzsymbols}
\newcommand*\colourcheckfire[1]{%
  \expandafter\newcommand\csname #1fire\endcsname{\textcolor{#1}{\Fire}}%
}
\colourcheckfire{red}

\definecolor{wcolor}{RGB}{103, 78, 167}

\definecolor{dcolor}{RGB}{166, 77,21}

	\definecolor{fastcolor}{RGB}{121,178,128}
	\definecolor{slowcolor}{RGB}{98,106,234}

	\definecolor{predictioncolor}{RGB}{0,255,0}
	\definecolor{labelcolor}{RGB}{255,0,0}

	\definecolor{demphcolor}{RGB}{144,144,144}

\newcommand{\app}{\raise.17ex\hbox{$\scriptstyle\sim$}}

\def\x{$\times$}
\newcolumntype{x}[1]{>{\centering\arraybackslash}p{#1pt}}

\newlength\savewidth

\definecolor{cadmiumgreen}{rgb}{0.0, 0.42, 0.24}

\definecolor{PineGreen}{rgb}{0.0, 0.47, 0.44}

\newcommand{\ourmodel}{\textsc{VMAs} }
\newcommand{\ourmodela}{\textsc{VMAs}}

\usepackage[capitalize]{cleveref}
\crefname{section}{Sec.}{Secs.}
\Crefname{section}{Section}{Sections}
\Crefname{table}{Table}{Tables}
\crefname{table}{Tab.}{Tabs.}

\def\confName{WACV}
\def\confYear{2025}

\begin{document}

\title{\ourmodela: Video-to-Music Generation via \\ Semantic Alignment in Web Music Videos}

\author{
Yan-Bo Lin$^{1}$\thanks{Work done during an internship at ByteDance Inc.}\quad\quad
Yu Tian$^{2}$\quad\quad
Linjie Yang$^{2}$\quad\quad 
Gedas Bertasius$^{1}$\quad\quad
Heng Wang$^{2}$
\\
$^{1}$UNC Chapel Hill\quad 
$^{2}$ByteDance Inc. 
\\
{\texttt{\{yblin,gedas\}@cs.unc.edu}}
\\
{\texttt{\{yutian.yt,linjie.yang, heng.wang\}@bytedance.com}}
}

\maketitle

\begin{abstract}
%
We present a framework for learning to generate background music from video inputs.
Unlike existing works that rely on symbolic musical annotations, which are limited in quantity and diversity, our method leverages large-scale web videos accompanied by background music. This enables our model to learn to generate realistic and diverse music. 
To accomplish this goal, we develop a generative video-music Transformer with a novel semantic video-music alignment scheme. Our model uses a joint autoregressive and contrastive learning objective, which encourages the generation of music aligned with high-level video content. We also introduce a novel video-beat alignment scheme to match the generated music beats with the low-level motions in the video. Lastly, to capture fine-grained visual cues in a video needed for realistic background music generation, we introduce a new temporal video encoder architecture, allowing us to efficiently process videos consisting of many densely sampled frames. 
We train our framework on our newly curated DISCO-MV dataset, consisting of 2.2M video-music samples, which is orders of magnitude larger than any prior datasets used for video music generation. Our method outperforms existing approaches on the DISCO-MV and MusicCaps datasets according to various music generation evaluation metrics, including human evaluation.
Results are available at \url{https://genjib.github.io/project_page/VMAs/index.html}

\end{abstract}  
\vspace{-5mm}
\section{Introduction}\label{sec:intro}
\vspace{\secmargin}
Enabled by modern technology, most people can create their own videos via platforms like TikTok, Instagram, and others. However, creating engaging videos often requires matching the video with the right background music, which is challenging and demands lots of time and expertise. In particular, adding background music to the video often involves a costly search for suitable music and aligning the patterns of pre-produced music with the particular video events. Nevertheless, a well-incorporated music soundtrack can significantly enhance the video viewing experience and make the video more engaging.
Thus, to reduce the manual effort for content creators, the last several years have witnessed an increased demand for automatic video background music generation methods~\cite{iccv23_v2music,arxiv23_suitable_video2music,acmmm21_cmt,aaai24_v2meow}.

\begin{table*}[t]
    \caption{
    \textbf{Comparison of different video-to-music generation datasets.} Our newly curated DISCO-MV dataset is orders of magnitude larger than the previous YouTube-8M subset~\cite{yt8m} and SymMV~\cite{iccv23_v2music} datasets, commonly used for music-to-video generation. Compared to other music generation datasets, our DISCO-MV is filtered to exclude videos with little visual variance and remove the vocals for improved background music quality. Since DISCO-MV is built from the existing publicly available datasets, it will be available under the same CC-BY-4.0 license. 
    }
    \vspace{\tabupmargin}
    \centering
    \setlength{\tabcolsep}{7pt} 
    \resizebox{0.99\textwidth}{!}{
    \begin{tabular}{l c c c c }
        \toprule
        Dataset  & SymMV & \makecell{YT-8M Subset}& DISCO-10M & DISCO-MV \\
       \toprule
       Number of Videos & 1.1K & 110K & 15.3M & 2.2M \\
       Avg. Frame Similarity   Within Video      & 0.64& 0.66    & 0.91                        & 0.61       \\
       Vocals Removed?     & N/A\protect\footnotemark  &\redxmark&\redxmark&\cadmiumgreencheck \\
       License &CC BY-NC-SA 4.0&CC-BY-4.0&CC-BY-4.0& CC-BY-4.0 \\
       \bottomrule
    \end{tabular}
    }
   \label{tab:dataset}
   \vspace{\tabmargin}
\end{table*}

Most prior video-based music generation approaches~\cite{iccv23_v2music,arxiv23_suitable_video2music,acmmm21_cmt} use symbolic music annotations (\eg MIDI), which store manually transcribed musical data in a digital format.
However, relying on such symbolic annotations for video-to-music generation is typically suboptimal for several reasons.
First, due to the limited expressivity of symbolic annotations, music generation methods trained on such data~\cite{iccv23_v2music,arxiv23_suitable_video2music,acmmm21_cmt} cannot capture subtle nuances in music soundtracks such as timbre variations, articulation, and the expressiveness of live performances. 
Moreover, the fidelity of the generated music is largely contingent upon the quality of a sound synthesizer or a MIDI playback engine, which may not adequately reflect the full depth and complexity of the actual instruments. 
Lastly, the small scale and limited genre diversity of MIDI annotations typically lead to poor generalization.

To address these issues, we propose to use a large collection of freely accessible internet videos that can provide rich supervision for learning an effective video-to-music generation model spanning many music genres.
Several recent methods~\cite{aaai24_v2meow,iclr23_discrete_music,icml23_long_video_music} have explored using Web videos for music generation. However, these methods are typically trained on small-scale datasets that span narrow domains (e.g., dancing videos).  
Instead, to leverage the scale and diversity of Web music videos, we develop a video-music generation framework, dubbed \ourmodela. Our approach leverages a generative video-music Transformer augmented with a novel \textbf{V}ideo-\textbf{M}usic \textbf{A}lignment \textbf{S}cheme (\ourmodela). We train our model using joint autoregressive and contrastive learning objectives. The generative objective focuses on realistic music generation, while the contrastive objective encourages alignment between generated music and \textit{high-level} video content (e.g., the genre of a video). We also introduce a novel video-beat alignment scheme that guides the generated music to be in harmony with \textit{low-level} visual cues, such as scene transitions or dynamic human motions.
Lastly, unlike existing video-music generation methods~\cite{iccv23_v2music,arxiv23_suitable_video2music,acmmm21_cmt,aaai24_v2meow} that use video encoders optimized for coarse spatial recognition~\cite{clip,i3d,vit,vit-vqgan}, we develop an efficient video encoder that can process videos with many densely sampled frames to capture fine-grained temporal cues needed for accurate music generation. 

We train our method on our newly curated DISCO-MV dataset, which consists of 2.2M video-music videos spanning 47,000 hours.
Compared to prior video-based music generation approaches, our method generates superior-quality music on the MusicCaps~\cite{arxiv23_musiclm} and  DISCO-MV test sets. We also conduct human evaluations and show that human subjects consistently prefer our generated music to music generated by previous methods on both music quality and music-video alignment.

\footnotetext{SymMV is in the MIDI format without vocal tracks.}


\section{Related Work}
\vspace{\secmargin}

\subsection{Text-Conditioned Audio and Music Generation}  
\vspace{\subsecmargin}
The emergence of powerful text models has led to a number of text-to-audio~\cite{arxiv23_audiobox,icml23_audioldm,arxiv23_audioldm2,taslp23_audiolm,icml23_make_an_audio,taslp23_diffsound,arxiv24_fastaudio} and text-to-music generation models~\cite{arxiv23_musiclm,musicgen,nips23_melody,arxiv23_Mousais,nips23_codi,arxiv23_codi2,iclr24_MAGNET}. %
Several recent models~\cite{audiogen,musicgen,taslp23_audiolm,arxiv23_musiclm,arxiv23_soundstorm} use a neural audio codec~\cite{tmlr23_encodec,taslp21_soundstream} first to convert a continuous audio signal into discrete tokens, and then train their conditional music generation model from textual prompts. %
Additionally, the recent diffusion-based audio and music generation methods~\cite{taslp23_diffsound,icml23_make_an_audio,icml23_audioldm,arxiv23_audioldm2,acmmm23_tango} convert raw audio into a spectrogram via a Short-Time Fourier Transform (STFT), and then use diffusion networks to predict a spectrogram from textual prompts. 
Despite the impressive progress of text-to-music generation models, text prompts can only provide high-level music characteristics such as genre and emotion, making it difficult to align the generated music with fine-grained video content. 
Instead, our work addresses this issue by generating music directly from the videos.  %

\subsection{Video-to-Music Generation} 
\vspace{\subsecmargin}

Many methods~\cite{arxiv23_foleygen,nips23_diff_foley,cvpr_owens2016visually,cvpr23_v2a,eccvw_chen2018visually,cvpr18_visual2sound,tip20_v2a,bmvc21_taming,cvpr23_physics,arxiv24_soundify} have tackled the general problem of video-to-audio generation in the recent years.
Building on this work, recent video-to-music generation methods have been designed to produce musical accompaniments for instrument performance in the video~\cite{gan2020foley,koepke2020sight,nips20_audeo}, to generate music that matches human motion/dance~\cite{nips19_dance2music,zhuang2022music2dance,shlizerman2018audio,eccv22_dnace2music,iclr23_discrete_music,icml23_long_video_music}, and also to synthesize background music for silent videos~\cite{acmmm21_cmt,iccv23_v2music,aaai24_v2meow,arxiv23_suitable_video2music}.
Most of these prior video-to-music generation approaches~\cite{iccv23_v2music,arxiv23_suitable_video2music,acmmm21_cmt} are trained using symbolic music annotations (\eg MIDI and ABC), which is often suboptimal because of the limited expressivity, diversity, and scale of such annotations.
Another line of relevant work involves music generation from dance videos~\cite{nips19_dance2music,zhuang2022music2dance,shlizerman2018audio,eccv22_dnace2music,iclr23_discrete_music,icml23_long_video_music}. However, as with symbolic music annotations, these methods are typically constrained by small and narrow training datasets. %
The recent work~\cite{aaai24_v2meow} tries to address these issues by using Internet videos~\cite{yt8m} to learn to generate music. However, this method also uses a small number of videos for training, leading to poor generalization. Instead, we propose a video-to-music generation framework, \ourmodela, that uses a novel video-music alignment scheme to match the generated music with the high-level and low-level visual cues, leading to vivid and realistic music generation. Furthermore, unlike all prior approaches, our proposed framework leverages orders of magnitude more videos, which is shown to be beneficial for both generation quality and video-music alignment. %

\begin{figure*}[t!]
    \centering
	\includegraphics[width=0.95\linewidth]{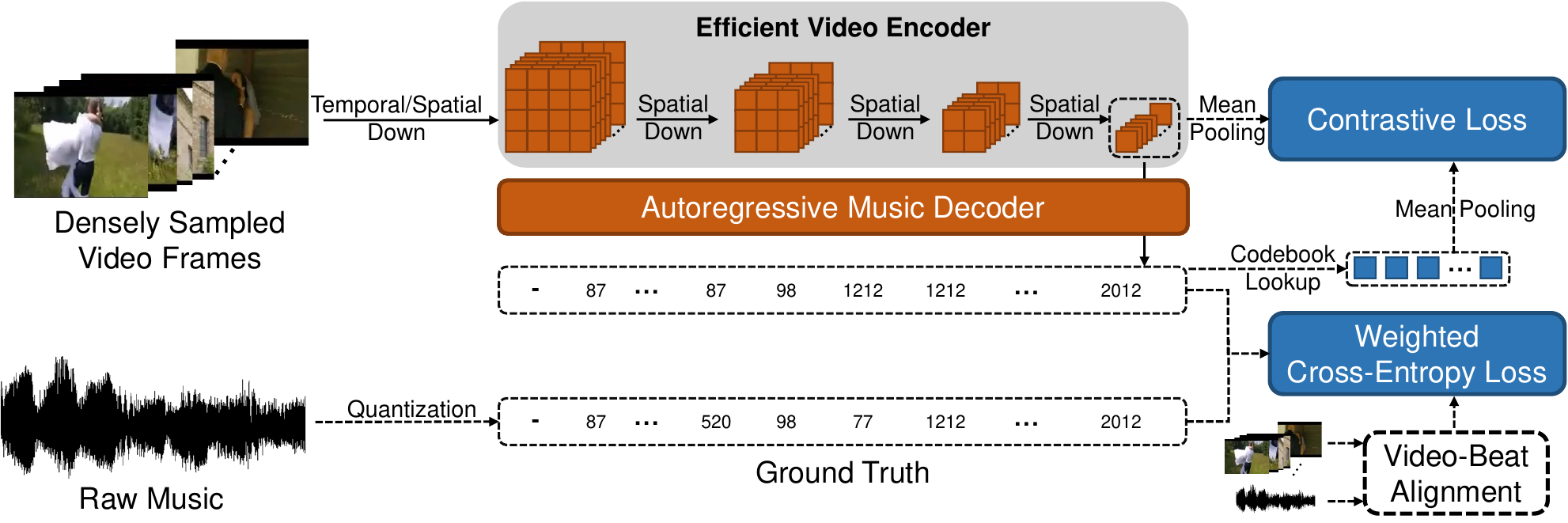}
     \caption{\textbf{Our Video-to-Music Generation Framework.} Our video-to-music generation framework consists of three main components: 1) an efficient video encoder for capturing fine-grained temporal cues from many densely sampled video frames, 2) an autoregressive music decoder for generating output audio tokens, and 3) a novel video-music alignment scheme that integrates a contrastive training objective and a novel video-beat alignment scheme, ensuring that the generated music exhibits \textit{high-level} and \textit{low-level} alignment with the video content.
    }
    \vspace{\figmargin}
	\label{fig:method}
\end{figure*}

\section{The DISCO-MV Dataset}
\vspace{\secmargin}

Prior video-to-music generation methods~\cite{acmmm21_cmt,iccv23_v2music,aaai24_v2meow,arxiv23_suitable_video2music} use relatively small training datasets, which limits their generalization. Instead, we believe that large-scale training data is the first requirement of building a highly performant video-to-music generation framework. Since the Web contains many freely available videos accompanied by music soundtracks, we aim to use such Web-based videos as a rich supervisory signal for training a large-scale video-to-music generation model. Using large-scale Web-based video-music data is also advantageous since the music soundtracks in Web videos are diverse and span various musical genres, enabling generalization to various scenarios.

To develop an expansive video-to-music generation dataset, we investigate using the recently introduced DISCO-10M~\cite{disco} dataset, which was created for music research and contains 15M videos collected via Spotify's artist graph~\cite{Spotify_fans_also_like}. 
The videos in DISCO-10M were collected by matching the metadata for artists' top tracks with the corresponding YouTube videos. 
Upon our preliminary analysis, we found that many videos in DISCO-10M have poor video-music correspondence, i.e., many of the videos consist of a single frame replicated over several minutes or hours or have very little visual variation, which makes them unsuitable for learning a high-quality video-to-music generation model.
To address this issue, we developed an automatic framework to remove such unsuitable videos. Specifically, for each video, we computed pairwise frame similarities within a video and only kept videos with an intra-video frame similarity below a certain threshold (\ie 0.7).
 
We also found that most videos in DISCO-10M contain music soundtracks with vocals, which is suboptimal for training video-to-music generation models~\cite{arxiv23_musiclm,arxiv23_noise2music,nips23_melody}.
To obtain training soundtracks with instrumental sounds only, we used Demucs~\cite{demucs} for vocal removal. Empirically, we found that we could successfully remove vocal sounds from approximately 95\% of the music soundtracks.

Ultimately, this led to 2.2M filtered videos from the original DISCO-10M dataset, which we also combined with 81K video-music samples from YouTube-8M. Our newly curated video-music generation dataset, DISCO-MV, contains 2.28M video-music samples spanning 47,000 hours. Furthermore, we also manually selected $1,120$ and $1,086$ videos with high-quality video-music correspondence to be used for validation and testing. In Table~\ref{tab:dataset}, we compare various characteristics of previously used video-music generation datasets. Our DISCO-MV is \textbf{2000}$\times$  larger than prior symbolic music annotation datasets such as SymMV. Furthermore, DISCO-MV has a much smaller average frame similarity than YouTube-8M-Music and DISCO-10M, demonstrating that our dataset features videos with rich visual content. Lastly, we note that since DISCO-MV is built using the publicly available video-music datasets (i.e., YouTube-8M~\cite{yt8m} and DISCO-10M~\cite{disco}), both of which are covered by the CC-BY-4.0 license, DISCO-MV will be available under this same license. %

%

%

\section{Technical Approach}
\vspace{\secmargin}
\label{sec:method}
In this section, we present our generative video-music framework. As illustrated in \figref{method}, our method is composed of three main components: 1) an efficient video encoder, 2) an autoregressive music decoder, and 3) a video-music alignment scheme that integrates a contrastive training objective and a novel video-beat alignment scheme for \textit{high-level} and \textit{low-level} video-music alignment, respectively. Below, we describe the details of each of these components.
%
%

\subsection{Audio and Video Inputs} 
\vspace{\subsecmargin}

\noindent\textbf{Audio Inputs.} Given a raw music waveform associated with the video, we first compute audio representation $\mathbf{X}^{(0)}_a \in \mathbb{R}^{t_a \times d}$, where the superscript $(0)$ indicates that this feature tensor will be used as an input to the first layer of our music decoder, while $t_a$ denotes the number of timesteps related to audio length.
Following MusicGen~\cite{musicgen}, we use EnCodec~\cite{tmlr23_encodec}, a state-of-the-art convolutional encoder-decoder framework, to transform continuous audio streams into a series of discrete tokens.
%
%
%
%
%
This procedure transforms the raw music waveform into the above-described quantized music tokens $\mathbf{X}^{(0)}_a$ 
%
associated with discrete labels, denoted by $\mathbf{Y} \in \mathbb{R}^{t_a \times c}$ with $c$ discrete categories. Each category represents a specific range of audio frequencies.
These discrete audio labels $\mathbf{Y}$ will be used as a supervisory signal for our video-to-music generation model.

\noindent\textbf{Video Inputs.} We consider video clips $\mathbf{V} \in \mathbb{R}^{t_v \times H \times W \times 3}$, each containing $t_v$ RGB frames with height $H$ and width $W$. We sample these frames uniformly from the entire input video. The video inputs are aligned with the corresponding audio segments.

%

\begin{table}[t]
    \centering
    \caption{
    \textbf{Our Efficient Video Encoder Architecture.} We modify the original Hiera architecture~\cite{icml23_hiera} to efficiently process videos consisting of many frames sampled at a high FPS rate. Specifically, we increase the strides of Q-Pooling layers to reduce spatial dimension in the early layers of the network while maintaining the temporal dimension throughout the network. Processing video frames at high temporal granularity allows our model to extract much more fine-grained temporal video cues than the original Hiera model needed for effective music generation.
    }
    \resizebox{0.99\linewidth}{!}{
        \begin{tabular}{c|c|c|c|c}
            \multicolumn{1}{c}{} & \multicolumn{2}{c}{Architecture Details} & \multicolumn{2}{c}{\makecell{Output Sizes $T$\x$S^2$\x D}} \\
            \hline
            Stage & Hiera~\cite{icml23_hiera} & Ours & Hiera~\cite{icml23_hiera} & Ours \\
            \hline
            Video Input & \multicolumn{2}{c|}{N/A} & \multicolumn{2}{c}{96\x3\x224$^\text{2}$} \\
            \hline
            \multirow{3}{*}{3D Conv}  & \multicolumn{2}{c|}{3\x7$^\text{2}$} &  \multirow{3}{*}{48\x56$^\text{2}$\x96 } &  \multirow{3}{*}{48\x56$^\text{2}$\x96 }\\
            & \multicolumn{2}{c|}{stride 2 $\times$ 4$^\text{2}$} & &  \\
            & \multicolumn{2}{c|}{padding 1$\times$ 3$^\text{2}$} & &\\
            \hline
            \makecell{Pooling Strides \\ at $2^{nd}$} & 1\x\hcolor{2}\x\hcolor{2} & 1\x\ocolor{4}\x\ocolor{4} & 48\x\hcolor{28}$^\text{2}$\x192 & 48\x\ocolor{14}$^\text{2}$\x192 \\
            \hline
            \makecell{Pooling Strides \\ at $5^{th}$} & 1\x\hcolor{2}\x\hcolor{2} & 1\x\ocolor{7}\x\ocolor{7} & 48\x\hcolor{14}$^\text{2}$\x384 & 48\x\ocolor{2}$^\text{2}$\x384 \\
            \hline
            \makecell{Pooling Strides \\ at $21^{st}$} & 1\x2\x2 & 1\x2\x2 & 48\x\hcolor{7}$^\text{2}$\x768 & 48\x\ocolor{1}$^\text{2}$\x768 \\
            \hline
            \makecell{Pooling Strides \\ at $24^{th}$} & 1\x1\x1 & 1\x1\x1 & 48\x\hcolor{7}$^\text{2}$\x768 & 48\x\ocolor{1}$^\text{2}$\x768 \\
            \hline
        \end{tabular}
    }
    \label{tab:hiera}
    \vspace{\tabmargin}
\end{table}

\subsection{Efficient Video Encoder}
\vspace{\subsecmargin}

Given an input video $\mathbf{V}$, we aim to compute a video representation that could be used for video-to-music generation.
%
%
Building on the developments in the video recognition domain, we use Hiera~\cite{icml23_hiera}, which demonstrated strong results on various video recognition tasks. 
While the original Hiera model was designed to operate on sparsely sampled video frames for coarse video recognition tasks, video-to-music generation requires fine-grained temporal information for accurate video-music alignment in time, such as matching music beats with the active motions in the video. To enable this capability, we modify the original Hiera architecture to allow it to efficiently process videos consisting of a much larger number of frames (\ie 96) sampled at a high FPS rate (\ie 9.6). %

Specifically, as depicted in \tabref{hiera}, we first use a 3D convolution with a $[3\times 7 \times 7]$ kernel and a $[2\times 4 \times 4]$ patch stride to obtain the visual tokens. 
%
Afterward, the original Hiera model performs three spatial downsampling operations, denoted as Q-Pooling, after the following transformer blocks: $2^{nd}$, $5^{th}$, and $21^{st}$.
To improve the efficiency of processing many frames sampled at high FPS, we increase the spatial downsampling rate of the first and second Q-Pooling layers from 2 to 4 and 2 to 7, respectively. This leads to feature tensors with a significantly reduced spatial resolution of $2\times2$ in the third stage and $1\times1$ in the final stage (assuming a $224\times224$ input size). 
We then directly obtain the final video representations $\mathbf{X}_v \in \mathbb{R}^{t_v/2 \times d}$,  where $t_v$ and $d$ represents the number of video frames and a channel dimension, respectively. %
These simple architecture modifications (see \tabref{hiera}) allow our video encoder to maintain high temporal resolution to capture fine-grained temporal video cues. %

%
%

 %

%

%
%

\subsection{Autoregressive Music Generation} 
\vspace{\subsecmargin}
\label{sec:music_generation}
We now describe our autoregressive music decoder that uses the video features $\textbf{X}_v$ and quantized music tokens $\mathbf{X}^{(0)}_a$ for music generation. 
Specifically, as our autoregressive music decoder, we adopt a standard transformer architecture with $L$ layers of alternating multi-head and causal attention blocks.
We feed the previously computed video features $\mathbf{X}_v$ into every multi-head attention as contextual features for autoregressive music generation. 
The operations in our decoder can be written as:
\begin{equation}
\begin{aligned}
\label{eq:spatial_att}
\mathbf{F}^{(\ell)} &= \mathrm{CA}(\mathbf{X}_a^{(\ell-1)}, \mathbf{X}_a^{(\ell-1)}, \mathbf{X}_a^{(\ell-1)}) + \mathbf{X}_a^{(\ell-1)} , \\
\mathbf{X}_a^{(\ell)} &= \mathrm{MHA}(\mathbf{F}^{(\ell)}, \mathbf{X}_v, \mathbf{X}_v) + \mathbf{F}^{(\ell)},
\end{aligned}
\vspace{\eqmargin}
\end{equation}
where $\mathrm{CA}(.)$ and $\mathrm{MHA}(.)$ are standard causal and multi-head attention blocks, and $\mathbf{F}^{(\ell)} \in \mathbb{R}^{t_a \times d}$ is an intermediate music representation computed from music tokens $\mathbf{X}^{(\ell-1)}_a$. %
The new music token representation $\mathbf{X}_a^{(\ell)}$ at layer $\ell$ is then computed using a multi-head self-attention that uses the intermediate music representation $\mathbf{F}^{(\ell)}$ as query, and video features $\mathbf{X}_v$ as keys and values.
The final music output $\mathbf{\hat{Y}} \in \mathbb{R}^{t_a \times c}$ is predicted via an MLP layer attached to the last decoder layer. %

%


\begin{figure*}[t!]
    \centering
	\includegraphics[width=0.99\linewidth]{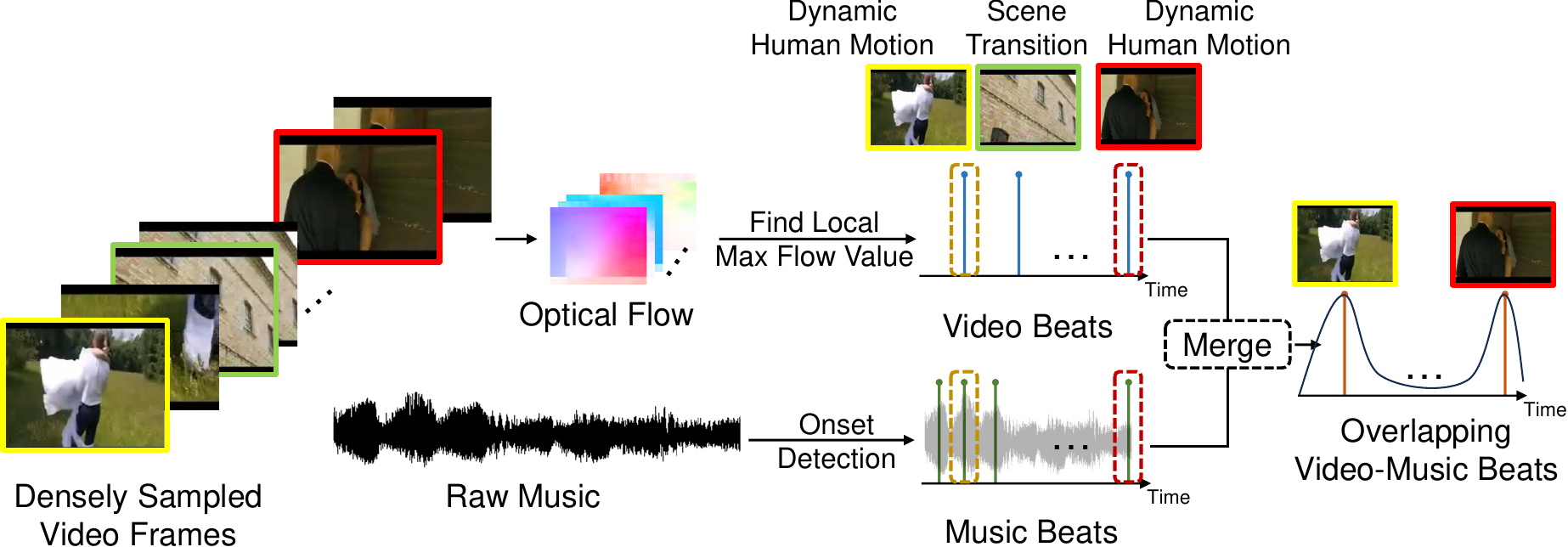}
     \caption{\textbf{Video-Beat Alignment Scheme.} 
    Our proposed alignment scheme allows us to detect moments in the video where music beats align with low-level visual cues such as dynamic human motions or scene transitions. We use Onset Detection~\cite{onset} and Optical Flow to identify such aligned video-beat moments. This information is then used to supervise our video-music generation model such that it would produce music aligned with low-level dynamic visual content.
    }
    \vspace{\figmargin}
	\label{fig:avbeat}
\end{figure*}

\subsection{Semantic Video-Music Alignment} 
\vspace{\subsecmargin}
To improve the alignment between generated music and video, we incorporate the following two strategies into our framework: (1) a global video-music contrastive objective, which guides the music to match the \textit{high-level} video cues (e.g.,  genre, style, etc.),
and (2) a fine-grained video-beat alignment scheme, designed to generate music beats that match the \textit{low-level} video cues such as scene transitions or dynamic human motions. We describe each of these schemes below.

\noindent\textbf{Global Video-Music Contrastive Objective.} 
To implement our contrastive video-music matching objective, we use the video representation $\mathbf{X}_v$ and compute the music features as $\mathbf{M} = \mathbf{\hat{Y}} \mathbf{E} $, where $\mathbf{\hat{Y}} \in \mathbb{R}^{t_a \times c}$ is the output from our music decoder, and $\mathbf{E} \in \mathbb{R}^{c \times d}$ is the embedding matrix from the EnCodec model. Conceptually, this operation allows us to map the discrete predicted audio categories $\mathbf{\hat{Y}}$ into higher-dimensional music embeddings for more effective contrastive matching.
To obtain an aggregated video representation  $\mathbf{\bar{X}}_{v} \in \mathbb{R}^{d}$, we apply temporal mean pooling on the video features $\textbf{X}_v$.
We also apply mean pooling on the music features $\mathbf{M}$ to generate an aggregated music representation $\mathbf{\bar{M}} \in \mathbb{R}^{d}$. %
Then, the video-music contrastive objective is written as:  
%
\begin{equation}
\begin{aligned}
\label{eq:contrastive}
\mathcal{L}_{c} = -\frac{1}{B} \sum_{i=1}^B {\rm log}  \frac{ {\mathrm exp} (\mathrm{g}(\mathbf{\bar{X}}_v(i),\mathbf{\bar{M}}(i)))} {\sum_{j=1}^{B} {\mathrm exp} (\mathrm{g}(\mathbf{\bar{X}}_v(i),\mathbf{\bar{M}}(j)))},
 \end{aligned}
\end{equation}

where $\mathrm{g}(\mathbf{x}, \mathbf{y})$ is the standard cosine similarity function and $B$ is the batch size.  %

\begin{table*}[t]
    \caption{
    \textbf{Comparison with the State-of-the-Art.} We evaluate all methods~\cite{acmmm21_cmt,arxiv23_suitable_video2music,musicgen,aaai24_v2meow} on MusicCaps~\cite{arxiv23_musiclm} and DISCO-MV using FAD, KL, and Music-Video Alignment (MV Align) metrics. 
    Since V2Meow's~\cite{aaai24_v2meow} code and pretrained models are unavailable, we only report their FAD results on MusicCaps.
    Our \ourmodel outperforms all prior approaches in all evaluation metrics on all datasets. 
    }
    \centering
    \setlength{\tabcolsep}{7pt}
    \resizebox{0.75\textwidth}{!}{
    \begin{tabular}{l ccc c c c }
        \toprule
        \multicolumn{1}{c}{}& \multicolumn{3}{c}{\footnotesize{MusicCaps} Test Set} & \multicolumn{3}{c}{\footnotesize{DISCO-MV}} 
        \\ \cmidrule{2-7}
        Method  & FAD $\downarrow$ & KL  $\downarrow$ & \makecell{MV Align  }$\uparrow$ & FAD $\downarrow$ & KL  $\downarrow$ & \makecell{MV Align  }$\uparrow$\\
       \toprule
        \multicolumn{7}{l}{\textit{\textbf{Symbolic Music}}}  \\
       CMT~\cite{acmmm21_cmt}                                                   & 16.2          & 1.42       & 0.18    & 3.70           & 1.82        & 0.34                   \\
       Video2Music~\cite{arxiv23_suitable_video2music}                          & 24.7          & 1.35      & 0.19     & 4.36           & 1.93        & 0.29             \\
       \midrule
       \multicolumn{7}{l}{\textit{\textbf{Waveform Music}}}  \\
       VidMusicGen~\cite{musicgen}                                              & 6.91          & 1.26       & 0.17    & 2.93           & 1.60        & 0.25             \\
       Vid2MLDM~\cite{acmmm23_tango}                                            & 8.99          & 1.15       & 0.20    & 3.21           & 1.41        & 0.32                         \\
       V2Meow\protect~\cite{aaai24_v2meow}                         & 4.62          & -          & -        & -                       & -        & -\\
      \midrule
      \multicolumn{7}{l}{\textit{\textbf{Waveform Music}}}  \\
       \ourmodel (Ours)                              & \bf4.07             & \bf1.09          & \bf0.22 & \bf2.38              & \bf1.34          & \bf0.35                   \\
       \bottomrule
    \end{tabular}
    }
   \label{tab:sota}
   \vspace{\tabmargin}
\end{table*}

\noindent\textbf{Video-Beat Alignment Scheme.} 
Compared to the high-level contrastive video-music matching objective above, our video-beat alignment scheme encourages low-level alignment between generated music and video dynamics (e.g., scene transitions, rhythmic human motions, etc.). 
Our proposed video-beat alignment scheme (shown in \figref{avbeat}) consists of several steps. First, we detect the music beats via off-the-shelf onset detection algorithm~\cite{onset}. 
Formally, we denote the detected music beats as $\mathbf{P}_a \in \mathbb{R}^{t_a}$, where $\mathbf{P}_a[t]$ is set to $1$ if a music beat was detected at timestep $t$; otherwise, it is set to $0$.
Afterward, we identify "video beats" by examining the magnitude of optical flow~\cite{raft} computed for all consecutive video frames in the video. %
We then spatially average the optical flow magnitudes across all pixels to compute the optical flow magnitude for each frame pair and store these values in $\mathbf{O} \in \mathbb{R}^{t_a}$ that is linearly interpolated to match the dimension of the music beats $\mathbf{P}_a$. 
Afterward, to obtain ``video beats'' $\mathbf{P}_v \in \mathbb{R}^{t_a}$, we set $\mathbf{P}_v[t]$ for each timestep $t$ to 1 when $\mathbf{O}[t]$ is the maximum in a temporal window of $\mathbf{O}[t-\delta:t+\delta]$, otherwise $\mathbf{P}_v[t] = 0$. Here, $\delta$ serves as a hyperparameter for controlling the size of a local temporal window, which determines the granularity of actions we aim to capture.
Conceptually, we aim to capture the timesteps corresponding to large optical flow values since they likely depict the dynamic video moments (e.g., human motions, scene changes, etc.).
Finally, we compute the overlap between the detected music and video beats as:
\vspace{-0.1cm}
\begin{equation}
\begin{aligned}
\label{eq:av_align}
\mathbf{P}_{av}[i]   = \begin{cases} 
    1 & \text{if}  \hspace{0.3em} \mathbf{P}_{v}[i]=1 \hspace{0.3em}  
       \text{and}  \hspace{0.3em} \sum_{j=-\delta}^{\delta} \mathbf{P}_{a}[i+j] > 0 \\
    \alpha & \text{else} \hspace{0.3em}. \\
    \end{cases}
\end{aligned}
\end{equation}

Intuitively, the formulation above computes the overlapping video-music beats by checking whether the video beats $\mathbf{P}_v[i]$ match the music beats in a window  $\mathbf{P}_a[i-\delta:i+\delta]$. 
We also introduce a hyperparameter $\alpha$ to prevent disproportionately de-emphasizing timesteps without the detected video-music beats during training. 
The detected video-music beats $\mathbf{P}_{av}$ are then incorporated into our autoregressive generation objective:

%

%
%
%
%
\begin{equation}
\begin{aligned}
\label{eq:att_celoss}
    \mathcal{L}_{g} = -\sum^{c}_{i=1} \mathbf{P}_{av} \mathbf{Y}_i \log(\hat{\mathbf{Y}}_i),
\end{aligned}
\end{equation}

where $\mathbf{Y}$ and $\hat{\mathbf{Y}}$ are the ground truth and predicted audio tokens, respectively. 
Intuitively, the detected video-music beats $\mathbf{P}_{av}$ are used as importance weights to guide our model to generate music beats aligned with low-level visual content.
Note that if all values in $\mathbf{P}_{av}$ are set to $1$, the \eqnref{att_celoss} represents the standard (i.e., uniformly weighted) autoregressive objective.
Our overall training objective is $\mathcal{L} = \beta \mathcal{L}_{c} + \mathcal{L}_{g}$, where $\beta$ is a balancing term between the contrastive and autoregressive terms.

\vspace{-0.2cm}
\section{Experimental Setup}
\vspace{\secmargin}
\label{sec:exp_setup}
\subsection{Downstream Datasets}
\vspace{\subsecmargin}
We train our approach on the DISCO-MV training set and evaluate on two video-music datasets test sets: MusicCaps~\cite{arxiv23_musiclm} and our DISCO-MV, which we describe below. 
\begin{itemize}
  \item \textbf{MusicCaps}~\cite{arxiv23_musiclm} evaluation set contains $2,858$ ten-second music video samples from YouTube paired with captions written by professional musicians.
  \item \textbf{DISCO-MV} is our newly curated dataset designed for video-to-music generation. We manually selected $1,120$ videos for validation and $1,086$ for testing, ensuring high-quality video-music correspondence in each set.
\end{itemize}
%
\subsection{Evaluation Metrics}
\label{sec:evaluation}
Following~\cite{aaai24_v2meow,musicgen,acmmm23_tango,nips23_melody}, we use a variety of metrics to evaluate our generated music samples:
\begin{compactitem}
    \item \textbf{Fréchet Audio Distance (FAD)}~\cite{fad} computes the distance between the distribution of the generated music and the reference music in the pretrained VGGish~\cite{vggish} feature space. 
    \item \textbf{KL Divergence (KL)} uses a music genre tagging model~\cite{music_tag_transformer} pretrained on the Million Song dataset~\cite{million_song} to measure the divergence of output distributions w.r.t the reference music. %

    \item \textbf{Music-Video Alignment (MV Align)}~\cite{aaai24_avalign} focuses on assessing the synchronization between music beats and visual dynamics in the video. %
    \item \textbf{Human Evaluation} asks the subjects to select the video-music samples they prefer based on 1) the overall music generation quality and 2) the alignment between the generated music and its corresponding video.
    %
    %
    Specifically, given a pair of video-music samples, where the video is the same but the music is generated by two different methods, the human raters are asked to select their preferred video-music sample based on the following prompts: 1) \textit{Which music video has higher overall quality music?} and 2) \textit{Which music video has better synchronization between music and visual content?}
    For each question, the subjects can choose one of the two methods or the third option ``Cannot tell."
    Each subject conducts 10 evaluations for a given pair of methods. The videos for each trial are randomly selected. The methods are unknown to the raters. 
    We compare our \ourmodel method to 4 competing approaches~\cite{acmmm21_cmt,arxiv23_suitable_video2music,musicgen,acmmm23_tango}. The performance is reported as the average human preference rate for each method. 
    We conduct our human study with 200 subjects on Amazon Mechanical Turk.
    %
    %
\end{compactitem}


\begin{figure*}[t!]
    \centering
	\includegraphics[width=0.99\linewidth]{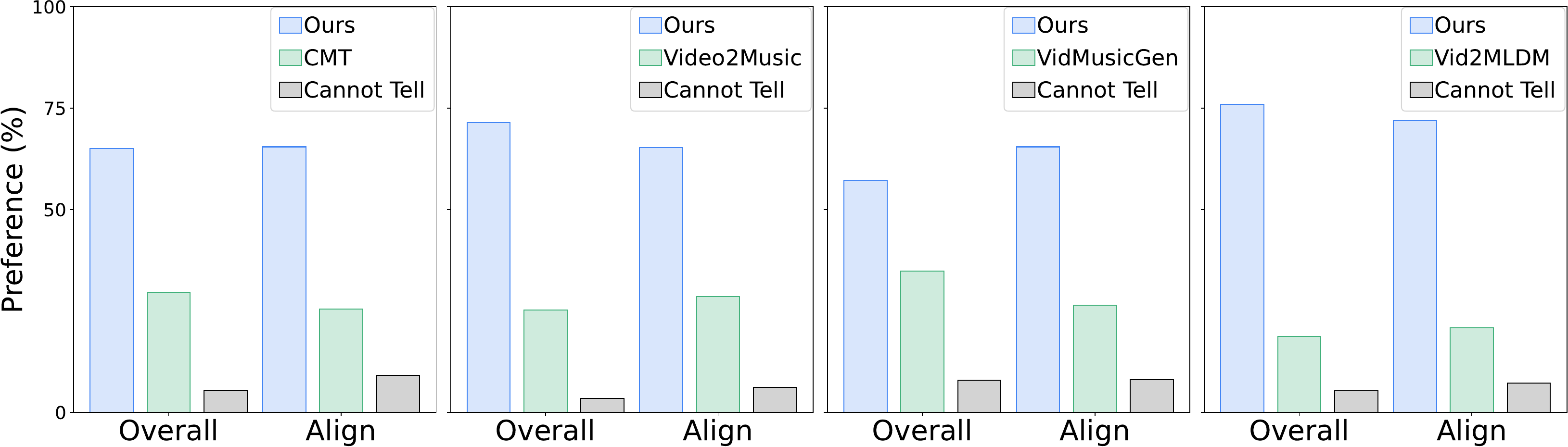}
     \caption{\textbf{Human Evaluation.} We conduct human evaluation to compare our \ourmodel method against several recent video-music generation methods~\cite{acmmm21_cmt,arxiv23_suitable_video2music,musicgen,aaai24_v2meow}. We present the results as average human preference ratings for 1) the overall music generation quality and 2) the alignment between generated music and the corresponding video content. Each comparison is conducted between a pair of methods. The methods are unknown to the human raters.
    Our results indicate that human subjects consistently prefer music-video samples with music generated by our method.
    }
    \vspace{\figmargin}
	\label{fig:sota_human_eval}
\end{figure*}

\subsection{Baselines}
\vspace{\subsecmargin}
We evaluate our model against the following recent video-music generation baselines. %
\begin{compactitem}
    \item \textbf{Controllable Music Transformer (CMT)}~\cite{acmmm21_cmt}. Given a video, this approach produces symbolic music outputs that are used for music generation. 
    Since this method relies on symbolic music annotations for training, it is not trainable on DISCO-MV. 
    Thus, we use the publicly available pretrained checkpoint to generate music on our two evaluation datasets.
    \item \textbf{Video2Music}~\cite{arxiv23_suitable_video2music}. Like CMT, Video2Music leverages semantic, motion, and emotional features to predict symbolic music outputs.
    Since it cannot be trained on raw music videos, we generated music by directly applying the pretrained model to the testing videos.
    
    \item \textbf{VidMusicGen}~\cite{musicgen}. We extend the text-to-music generation model MusicGen~\cite{musicgen} to video by using a pretrained music video captioning model~\cite{cap_model} to produce textual video descriptions, which are then fed to the pretrained MusicGen model for music generation.%
    \item \textbf{Vid2MLDM}~\cite{acmmm23_tango}. We adopt the pretrained text-to-music diffusion model, Tango \cite{acmmm23_tango}, for our video-to-music generation task by replacing its text encoder with our video encoder. 
    We then train Vid2MLDM using the same training and testing splits as our model.%
    \item \textbf{V2Meow}~\cite{aaai24_v2meow}. This method uses sparsely sampled video frames as inputs for video-to-music generation. It was trained on a subset of YouTube-8M~\cite{yt8m}. Due to the unavailability of source code or pretrained checkpoints, we directly report the results from their paper.
\end{compactitem}
\section{Results and Analysis}
\vspace{\secmargin}
\label{sec:results}
\subsection{Comparison with the State-of-the-Art}
\vspace{\subsecmargin}

In \tabref{sota}, we present the results of our \ourmodel method on the MusicCaps and DISCO-MV datasets. %
Our method outperforms all prior approaches in all evaluation metrics on all datasets, indicating the superiority of our proposed method compared to the existing video-to-music generation approaches. We also make several other interesting observations.
First, we note that the symbolic music generation approaches, CMT and Video2Music, obtain inferior results compared to other approaches (including ours). This confirms our earlier hypothesis that the approaches relying on symbolic annotations suffer from small amounts of training data and limited diversity in the annotations.
Second, we demonstrate that our autoregressive approach achieves better results than the diffusion-based Vid2MLDM~\cite{acmmm23_tango} in all metrics on all datasets.
Third, we show that our model outperforms the state-of-the-art V2Meow on MusicCaps. Since V2Meow's source code has not been publicly released, we can only perform comparisons using the results reported in their paper.
For a fair comparison with V2Meow, in \tabref{v2meow}, we also include the variant of our \ourmodel model trained using the same data as V2Meow.
Our results indicate that our \ourmodel surpasses V2Meow even when both methods are trained on the same data. This indicates the effectiveness of our architecture design and also the importance of our video-music alignment training schemes. 
We also observe that by using our DISCO-MV dataset, we can improve our model's performance substantially, thus demonstrating the importance of our large-scale dataset. %

\begin{table}[t]
    \caption{
    \textbf{Comparison to V2Meow Using the Same Training Data.} For a fair comparison with V2Meow, we include the variant of our \ourmodel model trained using the same data as V2Meow. Both methods are evaluated on MusicCaps using the FAD and KL Div. computed via Leaf classifer~\cite{leaf}. 
    }
    \vspace{\tabupmargin}
    \centering
    \setlength{\tabcolsep}{7pt}
    \resizebox{0.49\textwidth}{!}{
    \begin{tabular}{l cccc }
        \toprule
        Method & \makecell{Training  Dataset}  & \#Videos & FAD $\downarrow$ & KL. $\downarrow$\\
       \toprule
       V2Meow~\cite{aaai24_v2meow}           & MV100K           & 100K        & 4.62 & 1.22                 \\
       \ourmodel (Ours)                            & MV100K           & 100K        & 4.51   & 1.15               \\
       \ourmodel (Ours)                             & DISCO-MV    & 2.2M        & \textbf{4.07} & \bf1.10                  \\
       \bottomrule
    \end{tabular}
    }
   \label{tab:v2meow}
   \vspace{-5 mm}
\end{table}

\textbf{Human Evaluation.} In \figref{sota_human_eval}, we report the results of our human evaluation. We use two metrics: 1) overall music generation quality
and 2) alignment between generated music and the corresponding video content.
These results are presented as human preference rates, quantifying the percentage that the subjects favor our method over a competing method. All comparisons are made between a pair of methods, e.g., our \ourmodel vs. CMT, Video2Music, VidMusicGen, and Vid2MLDM. 
From the results in \figref{sota_human_eval}, we observe that human subjects consistently prefer our method to all the other methods according to both evaluation metrics.
Specifically, on average, our method is preferred over \textbf{70\%} of the time for overall music generation quality and \textbf{67\%} for the metric measuring the alignment between generated music and the corresponding video content.
%
%

%
%

%

\subsection{Ablation Studies}
\vspace{\subsecmargin}

\begin{table}[t]
    \centering
    \caption{\textbf{Effectiveness of Video-Music Alignment Schemes.} 
    Here, we study the impact of our two video-music alignment schemes: 1) video-music contrastive objective and 2) video-beat alignment. 
    We observe that using both alignment schemes allows us to obtain the best performance in FAD, KL, and MV Alignment metrics on DISCO-MV.
    }
    \vspace{\tabupmargin}
    \setlength{\tabcolsep}{7pt}
    \resizebox{0.49\textwidth}{!}{
    \begin{tabular}{lccc}
        \toprule
        Configuration & FAD $\downarrow$ & KL $\downarrow$ & MV Align $\uparrow$ \\
        \midrule
        Autoregressive Baseline            & 2.75 & 1.40 & 0.243 \\
        + Video-Music Contrastive        & 2.40 & \bf 1.34 & 0.251 \\
        + Video-Beat Alignment           & \bf2.38 & \bf1.34 & \bf0.352 \\
        \bottomrule
    \end{tabular}}
    \label{tab:abs}
    \vspace{\tabmargin}
\end{table}

\textbf{Importance of Video-Music Alignment.} In \tabref{abs}, we evaluate the importance of our two video-music alignment schemes: 1) video-music contrastive objective and 2) video-beat alignment scheme. 
Based on these results, we observe that the contrastive video-music matching objective produces a significant improvement in FAD (from \textbf{2.75} to \textbf{2.40}) and a significant improvement in KL (from \textbf{1.40} to \textbf{1.34}) metrics.
Furthermore, our proposed video-beat alignment scheme significantly improves the music-video alignment metric (\textbf{+0.109}).
By integrating these two complementary video-music alignment schemes, our model achieves optimal performance according to all three metrics.%

\textbf{Comparison to Other Video Encoders.} In \tabref{fps}, we compare our efficient video encoder with CLIP~\cite{clip} and Hiera~\cite{icml23_hiera} on DISCO-MV. All comparisons are done by incorporating each encoder into our video-to-music generation pipeline. %
Despite using a larger number of video frames (\ie  96 vs. 16), our video encoder is more computationally efficient than either CLIP or Hiera (\ie \textbf{130.7} vs. \textbf{140.2} and \textbf{281.6} in GFLOPs) and it achieves better music generation results. 

\textbf{Impact of Training Data Size.}
We next study the performance of our model when training it with different fractions of DISCO-MV data (i.e., 10\%, 25\%, 50\%, 100\%). We report the FAD results on MusicCaps and DISCO-MV test sets. 
On both datasets, we observe a consistent improvement with an increasing amount of training data (\ie MusicCaps: \textbf{4.7}, \textbf{4.4}, \textbf{4.3}, \textbf{4.1} and DISCO-MV: \textbf{3.2}, \textbf{2.9}, \textbf{2.7}, \textbf{2.4}), indicating the importance of large-scale training data for music generation. More results are shown in Appendix.

\textbf{Imapct of Data Cleaning.}
We next study the importance of our data cleaning protocol, which removes videos from DISCO-10M~\cite{disco} with high intra-frame similarity (i.e., videos with low visual variance). %
We train two variants of \ourmodel: 1) a variant trained on the full unfiltered DISCO-10M dataset vs. 2) a variant trained on our filtered DISCO-MV dataset. 
We report that training \ourmodel on the full DISCO-10M only achieves an FAD of \textbf{4.52}, a KL of \textbf{2.11}, and an MV-Align score of \textbf{0.18}. 
In contrast, training \ourmodel on our filtered DISCO-MV leads to a substantially improved performance across all metrics, achieving FAD of \textbf{2.38}, a KL of \textbf{1.34}, and an MV-Align score of \textbf{0.35}.
The results suggest that filtering the noisy videos in DISCO-10M is necessary for better video-to-music generation performance.

\begin{table}[t]
    \caption{
    \textbf{Comparison to Other Video Encoders.} 
    We compare the effectiveness of our proposed video encoder architecture to several existing video encoders~\cite{clip,icml23_hiera} for the video-to-music generation task. We evaluate the results on the DISCO-MV dataset using FAD, KL, and MV Align metrics. Based on these results, we observe that our proposed video encoder architecture is not only more efficient but also leads to higher-quality music generation, justifying our model design. 
    }
    \vspace{\tabupmargin}
    \centering
    \setlength{\tabcolsep}{7pt}
    \resizebox{0.49\textwidth}{!}{
    \begin{tabular}{l c c c c c}
        \toprule
        \makecell{Video \\ Encoder} & \#Frames & FAD$ \downarrow$ & KL  $\downarrow$ & \makecell{MV \\Align} $\uparrow$ & \makecell{GFLOPS  $\downarrow$} \\
       \midrule
       CLIP~\cite{clip}      & 16 & 2.61            & 1.41        &0.274& 281.6        \\
       Hiera~\cite{icml23_hiera}     & 16 & 2.58            & 1.41        & 0.316 &140.2        \\
       Ours       & 96           & \bf 2.38              & \bf 1.34          & \bf 0.342   & \bf130.7    \\
       \bottomrule
    \end{tabular}
    }
   \label{tab:fps}
   \vspace{\tabmargin}
\end{table}

\begin{figure*}[t!]
    \centering
	\includegraphics[width=0.85\linewidth]{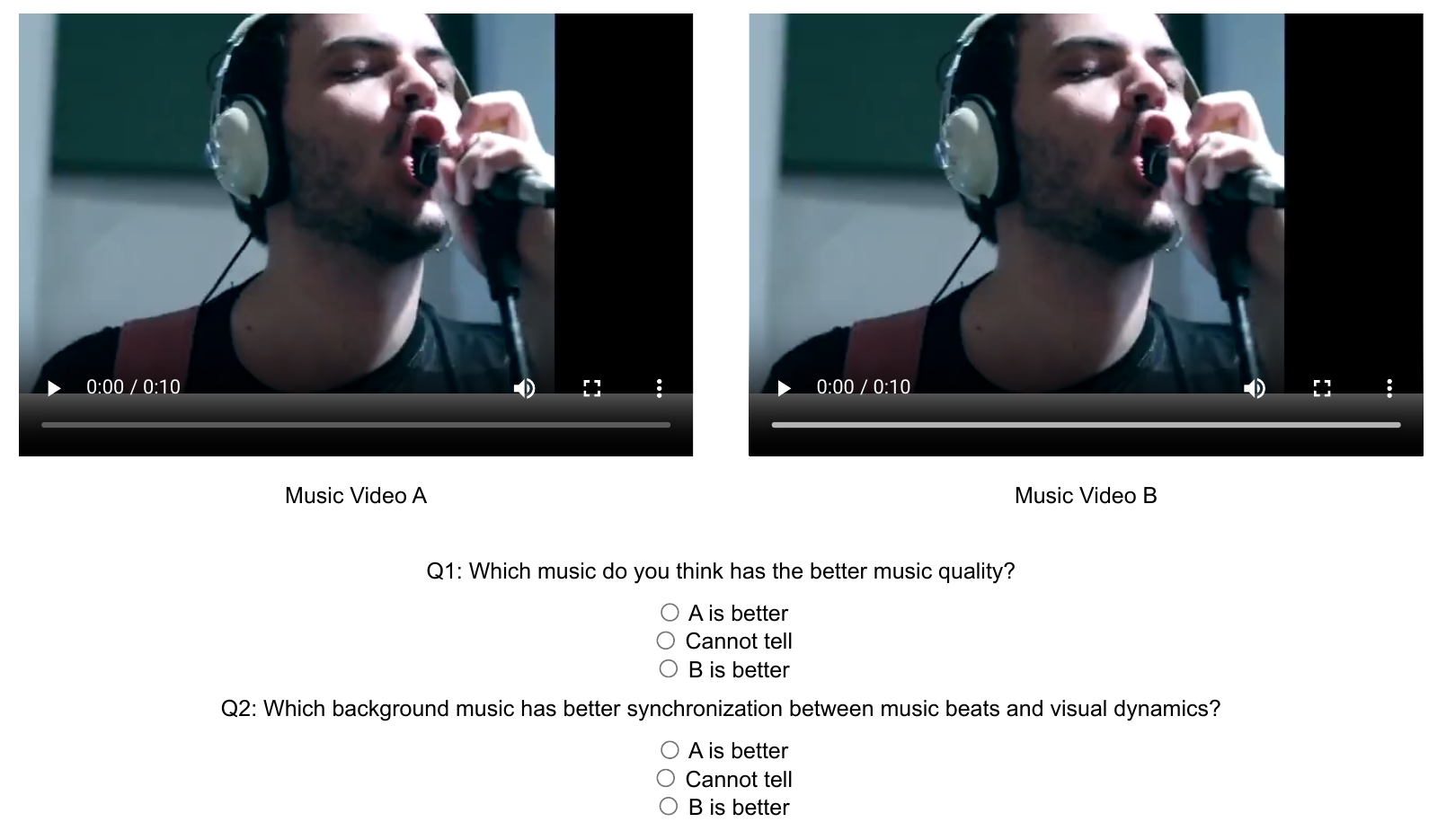}
     \caption{\textbf{Human Evaluation.} Human raters are asked to select the generated music that best aligns with a given video and the best music quality. We report the average human preference rate for each method. Note that all samples are present in a random order.
    }
    \vspace{\figmargin}
	\label{fig:human}
\end{figure*}

\section{Discussion and Conclusions}
\vspace{\secmargin}
\label{sec:conclusions}
In this paper, we presented a large-scale video-music generation framework that leverages our newly curated DISCO-MV dataset, which is orders of magnitude larger than any existing video-music datasets. 
Our proposed framework leverages a novel video-music alignment scheme consisting of 1) contrastive video-music matching and 2) video-beat alignment objectives to align the generated music with the high-level (e.g., genre) and low-level (e.g., scene transitions) visual cues in the video.
Furthermore, to effectively learn fine-grained visual cues from long, densely sampled video inputs, we designed a new video encoder, which we showed to be both more effective and more efficient than the existing encoders for the video-music generation task.
In the future, we will extend our framework to other video-music modeling problems such as video-music style transfer, video-music editing, and video-music retrieval. We will also investigate a joint video-and-text-to-music generation architecture that provides users with more control when generating music.

\section*{Acknowledgments}
 We thank Feng Cheng, Md Mohaiminul Islam, Ce Zhang, and Yue Yang for their helpful discussions. This work was supported by the Sony Faculty Innovation Award, Laboratory for Analytic Sciences via NC State University, ONR Award N00014-23-1-2356.

\newcount\cvprrulercount
\appendix
\section{Appendix Overview}
Our appendix consists of:

\begin{enumerate}
    \item Implementation Details.
    \item Human Evaluation Details.
    \item Music Genre Analysis of DISCO-MV 
    \item Additional Quantitative Results.
    \item Qualitative Results.
    \item A Supplementary Video. 
\end{enumerate}

\begin{figure*}[t!]
    \centering
	\includegraphics[width=0.82\linewidth]{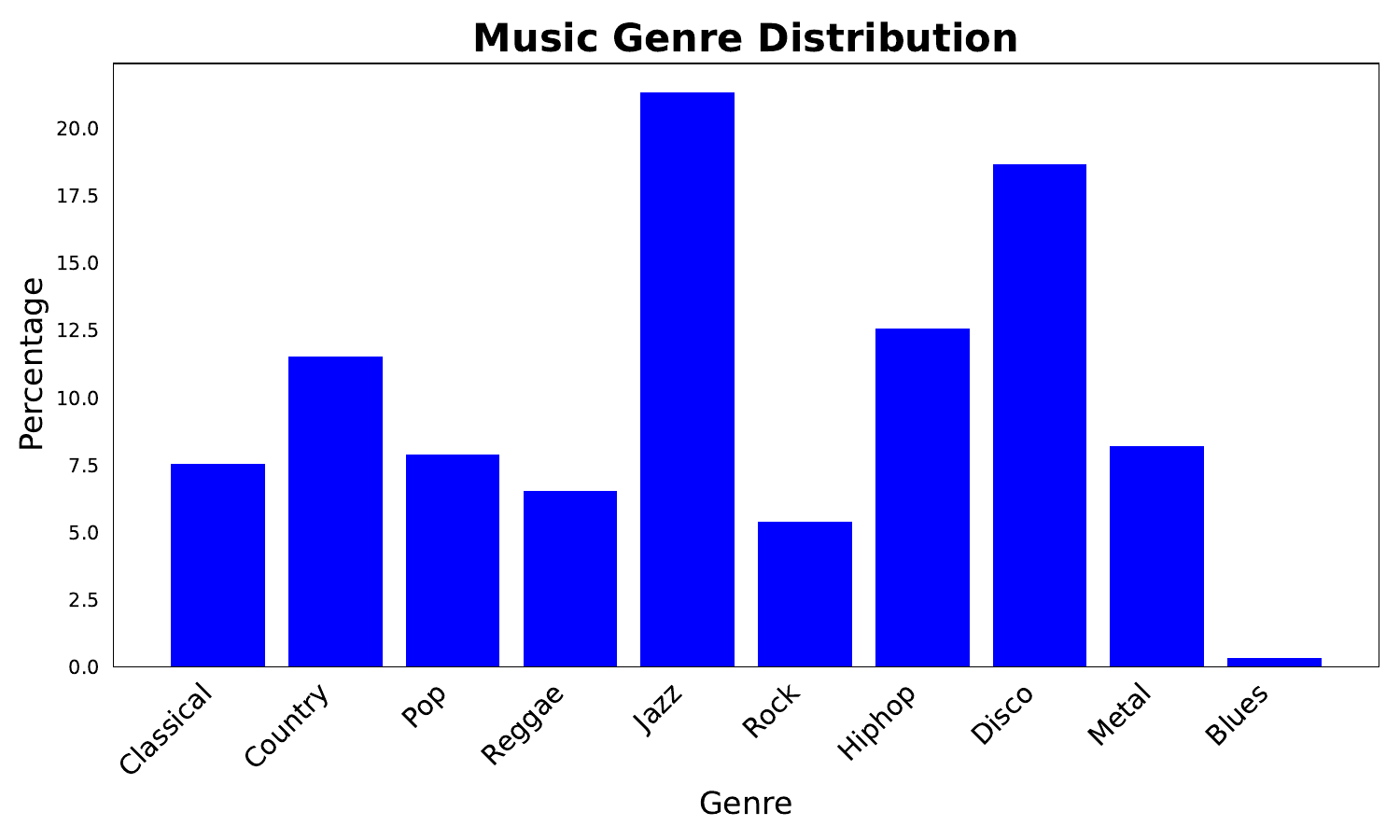}
     \caption{\textbf{Music Genre Distribution.} We present the GTZAN genres~\cite{gtzan} for the DISCO-MV dataset. Genres are assigned to each soundtrack based on the maximum cosine similarity between its sound embedding and the corresponding genre (text) embedding.
    }
    \vspace{\figmargin}
	\label{fig:disco-mv}
\end{figure*}

\section{Implementation Details}
\noindent\textbf{Audio Tokenization Model and Patterns.}
To transform a continuous $32$ kHz audio into discrete audio tokens, we leverage the pretrained EnCodec~\cite{tmlr23_encodec} with a stride of $640$,
resulting in a frame rate of $50$ Hz and an initial hidden feature size of $64$.  
The embeddings are quantized using a Residual Vector Quantization (RVQ) with four quantizers, each having a codebook size of $2048$.
As for the codebook pattern, we adopt the delay interleaving pattern~\cite{musicgen} to translate $10$ seconds of audio into $500$ autoregressive steps (audio tokens). 

\noindent\textbf{Efficient Video Encoder.}
Given an input video, we extract frames at a $9.6$ FPS rate, resulting in a total of $96$ video frames with a resolution of $224 \times 224$ for a $10$-second clip.
We utilize the pretrained Hiera-Base model \cite{icml23_hiera}, which consists of $24$ layers and performs downsampling three times through pooling.
These $96$ video frames are initially processed by a 3D CNN with a $[3 \times 7 \times 7]$ kernel, a $[2 \times 4 \times 4]$ stride, and a $[1 \times 3 \times 3]$ padding for video tokenization.
Subsequently, we adjust the original spatial downsampling method in Hiera, namely Q-Pooling, which utilizes the same size of pooling kernels and strides.
The Q-Pooling kernel sizes for the first and second downsampling stages are increased from $2$ to $4$ (at the $2^{\text{nd}}$ layer) and from $2$ to $7$ (at the $5^{\text{th}}$ layer), respectively.
For the third downsampling stage, the Q-Pooling kernel size is maintained at $2 \times 2$, implemented at the $21^{\text{st}}$ layer.
At the $24^{\text{th}}$ layer, the final layer, the model only increases the channel dimension through the liner projection, yielding video representations of dimension $48 \times 1 \times 1 \times 768$.

\begin{figure}[t!]
    \centering
    \includegraphics[width=0.45\textwidth]{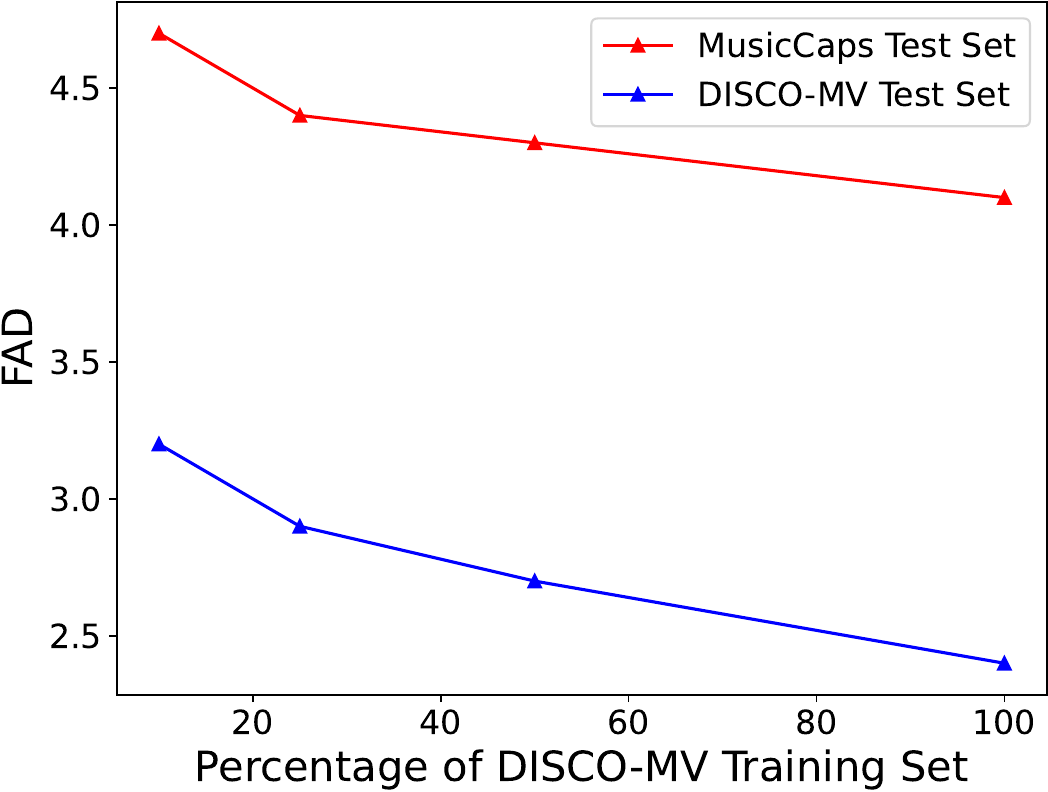}
    \caption{
        \textbf{Impact of the Training Data Size.} We train our method with increasing subsets of DISCO-MV and then evaluate on MusicCaps (\textbf{\textcolor{red}{Red}}) and  DISCO-MV (\textbf{\textcolor{blue}{Blue}}) using FAD metric (the lower the better). Our results illustrate that the size of video-music training data has a significant impact on the generated music quality. These results justify our approach of using Web videos for scaling our video-music training data.
    }
    \label{fig:scale_fad}
\end{figure}

\begin{figure}[t!]
    \centering
	\includegraphics[width=0.45\textwidth]{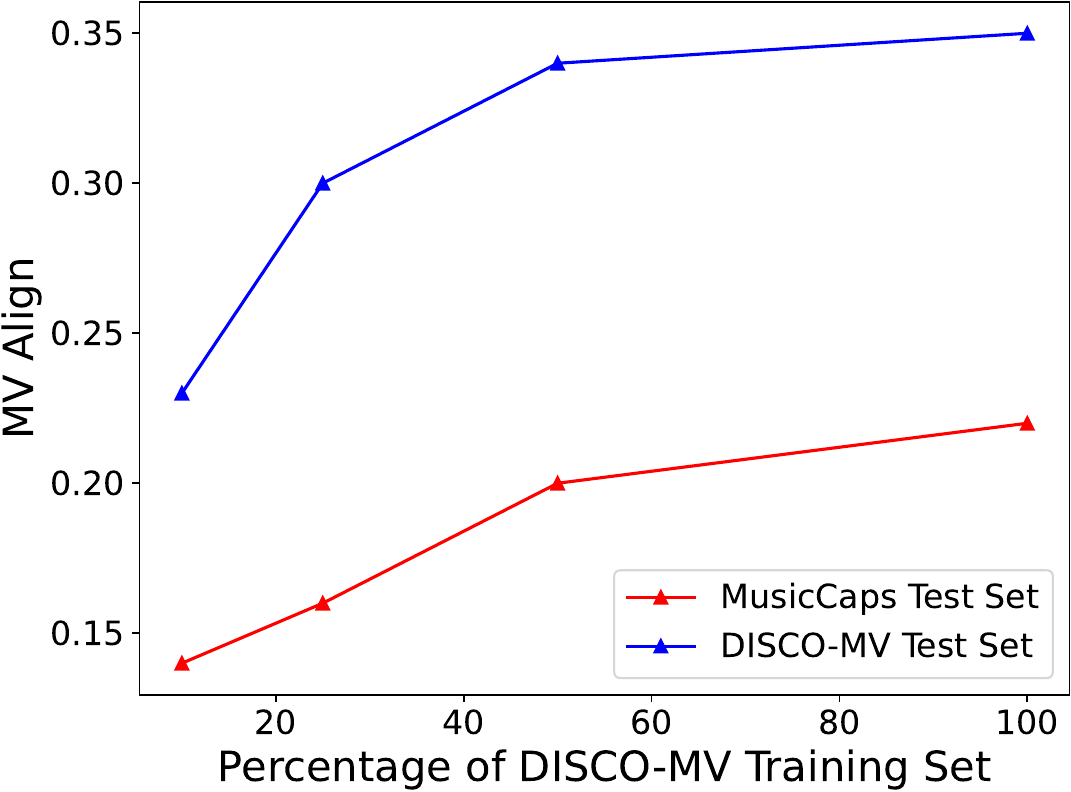}
     \caption{\textbf{Impact of the Training Data Size.} We train \ourmodel with increasing subsets of DISCO-MV and then report the Music-Video Alignment (MV Align) metric on MusicCaps (\textbf{\textcolor{red}{Red}}) and  DISCO-MV (\textbf{\textcolor{blue}{Blue}}).
    }
    \vspace{\figmargin}
	\label{fig:mvalign}
\end{figure}
\noindent\textbf{Autoregressive Audio Decoder.}
We adopt the autoregressive transformer models of pretrained MusicGen-medium~\cite{musicgen} as our autoregressive audio decoder.
The decoder consists of $48$ transformer layers in a feature dimension of $1536$  with $24$ standard causal and multi-head attention blocks. 

\begin{table*}[t]
    \caption{
    \textbf{Video-to-Music Retrieval Results.} We conduct a comparison of our designed efficient video encoder with existing video encoders~\cite{clip,icml23_hiera} for video-to-music retrieval. Our efficient video encoder is generalizable to the video-to-music retrieval task.
    }
    \centering
    \setlength{\tabcolsep}{6pt}
    \resizebox{0.75\textwidth}{!}{
    \begin{tabular}{l c c c c c c c}
        \toprule
        Method &\makecell{Video Encoder} & \#Frames & \makecell{V2M R@1  $\uparrow$} & \makecell{V2M R@10  $\uparrow$}\\
       \toprule
       \multirow{2}{*}{MVPt~\cite{cvpr22_mv_retrieval}}&CLIP~\cite{clip}      & 16 & 4.2       &  17.3 \\
                                                        &Hiera~\cite{icml23_hiera}     & 16  &4.8        & 18.1 \\
       \ourmodel &\ourmodel       & 96            & \bf6.3    & \bf26.7\\
       \bottomrule
    \end{tabular}
    }
   \label{tab:retrieval}
   \vspace{\tabmargin}
\end{table*}

\noindent\textbf{Optimiazation.}
We train \ourmodel on $10$-second video clips from DISCO-MV using the AdamW optimizer \cite{adamw} with a batch size of $8$ on each GPU.
The training takes approximately four days using 32 NVIDIA GPUs across 4 nodes. Each node is equipped with 8 GPUs, 92 CPUs, and 1000G of memory.
We utilize D-Adaptation~\cite{d-adaptation} to select the overall learning rate automatically. 
A cosine learning rate schedule with a warmup of $4000$ steps is deployed alongside an exponential moving average with a decay of $0.99$.
We set $\alpha$ in eq(3) to $0.05$ and  $\beta$ in the main draft to $0.25$. 

\section{Human Evaluation Details}
As depicted in \figref{human}, given a pair of video-music samples with the same video but different music generated by two methods, human raters are asked to choose their preferred video-music sample based on the following prompts: 1) \textit{Which music do you think has the better music quality?} and 2) \textit{ Which background music has better synchronization between music beats and visual dynamics?"} 
For each question, the subjects can choose one of the two methods or the third option "Cannot tell."
We collected approximately 200 subjects in the human evaluation by presenting results from a random method against \ourmodela.
In each survey, Each conducts 10 evaluations randomly selected among 50 evaluations.

\section{Music Genre Analysis of DISCO-MV}
Following the setup in DISCO-10M~\cite{disco}, we implement zero-shot music genre classification for DISCO-MV by utilizing pretrained CLAP~\cite{clap} embeddings extracted from 10-second music clips. 
Genre classification is conducted through genre-specific prompts ("This audio is a \textless genre\textgreater\ song") and identifying the genre via top-1 cosine similarity in a shared latent space for each music clip.
In \figref{disco-mv}, we report the GTZAN genre distribution~\cite{gtzan} of our DISCO-MV dataset. 
Jazz and disco genres are predominant while ensuring a wide range of musical diversity. 
However, we note that the blues genre has few samples in DISCO-MV due to the limited number of blues music in the original DISCO-10M dataset.

\section{Additional Quantitative Results}
\noindent\textbf{Impact of Training Data Size.} Similar to our analysis in the main draft, in \figref{scale_fad} and \figref{mvalign}, we visualize our model's performance on DISCO-MV as the training data increases in the FAD and Music-Video Alignment (MV Align) metrics, respectively. 
We observe consistent improvement in both metrics when the DISCO-MV data size increases. 
Despite the Vid2MLDM model not accounting for low-level music-video beat synchronization, its performance improves with more training data.
These additional results confirm that our large-scale DISCO-MV can improve music-video beat alignment as well as music quality with larger training data scales.

\textbf{Video-Music Retrieval Task.}
To evaluate our approach's generalization capability, in \tabref{retrieval}, we also compare \ourmodel against MVPt~\cite{cvpr22_mv_retrieval} for the video-to-music retrieval tasks on DISCO-MV.
Following the pipeline and size evaluation set of video-to-music retrieval framework~\cite{cvpr22_mv_retrieval}, we randomly sample 2,000 videos in the DISCO-MV test set and extracted music and video representations using our trained model on video-to-music generation task.
We used these feature representations for the video-to-music retrieval task and measured performance using the standard Recall@1 and Recall@10 metrics. 
For a fair comparison, we implement MVPt~\cite{cvpr22_mv_retrieval} using CLIP~\cite{clip} and Hiera~\cite{icml23_hiera} as the video encoders and train it under the same conditions as \ourmodela.
These results indicate that  \ourmodel is more accurate and generalizes better compared to MVPt (\ie \textbf{26.7} vs. \textbf{18.1} and \textbf{17.3} R@10) on the video-to-music task.

%
%

\begin{figure*}[t!]
    \centering
	\includegraphics[width=0.99\linewidth]{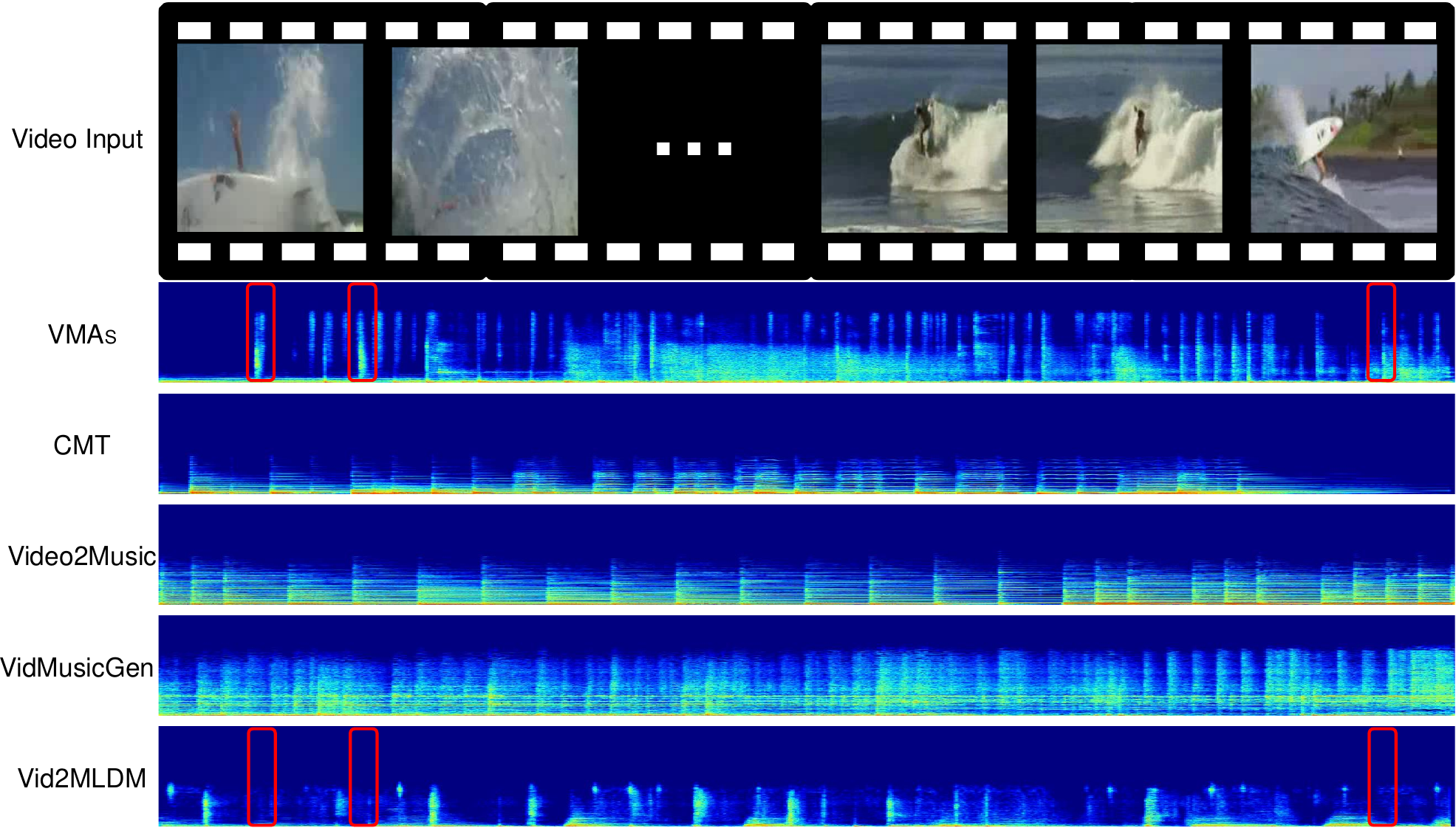}
     \caption{\textbf{Qualitative Video-to-Music Generation Results.}
        Here, we illustrate qualitative music generation results for a given silent video input. The generated music sample is visualized as a spectrogram.
        We compare the results of our model with CMT~\cite{acmmm21_cmt}, Video2Music~\cite{arxiv23_suitable_video2music}), VidMusicGen~\cite{musicgen} and Vid2MLDM~\cite{acmmm23_tango}.
        We note that most prior video-to-music generation approaches produce music beats of uniform intensity. In contrast, our model generates music beats that align well with dynamic video content, i.e., significant movements when a surfer changes direction during a sharp turn in this particular example.
    }
    \vspace{\figmargin}
	\label{fig:vis}
\end{figure*}

\section{Qualitative Results}
In \figref{vis}, we visualize our generated music results as a 2D spectrogram. We also include the results of the following video-to-music generation methods: CMT~\cite{acmmm21_cmt}, Video2Music~\cite{arxiv23_suitable_video2music}, VidMusicGen~\cite{musicgen} and Vid2MLDM~\cite{acmmm23_tango}. All results are obtained using the same video input shown at the top of the Figure.

Based on these results, we observe that symbolic music generation methods(\ie CMT and Video2Music) often generate music with uniform music beat patterns, which is suboptimal as the music fails to match the temporal dynamics of the video content. 
Furthermore, the existing waveform methods (\ie VidMusicGen and Vid2MLDM) struggle to generate music consistently synchronized with dynamic low-level video events.%
In comparison, the music beats generated by our model (highlighted in red boxes) have higher intensity when a surfer in the video performs a dramatic turn. This demonstrates that our model generates music that reflects the pace and magnitude of the actions occurring in the video. %
%
%

%

{
\bibliographystyle{ieee_fullname}
\bibliography{main}
}


\end{document}